\documentclass{article}
\usepackage{geometry}
\usepackage{appendix}
\usepackage{authblk}
\usepackage{amsfonts}
\usepackage{amsmath}
\usepackage{amssymb}
\usepackage{caption}
\usepackage{graphicx,subfig}
\usepackage{float}
\usepackage{algorithm}
\usepackage{algorithmicx}
\usepackage{algpseudocode}
\usepackage{amsmath}
\usepackage{latexsym}
\usepackage{multirow}

\author[1]{Fuzhou Gong \thanks{Corresponding author: fzgong@amt.ac.cn}}
\author[1]{Zigeng Xia \thanks{Corresponding author: xiazigeng@amss.ac.cn}}
\affil[1]{Academy of Mathematics and Systems Science, Chinese Academy of Sciences}
\title{Interpreting Deep Learning by Establishing a Rigorous Corresponding Relationship with Renormalization Group
}
\begin{document}
	\maketitle
\begin{abstract}
\setlength\parindent{1em}In this paper, we focus on the interpretability of deep neural network. Our work is motivated by the renormalization group (RG) in statistical mechanics. RG plays the role of a bridge connecting microscopical properties and macroscopic properties, the coarse graining procedure of it is quite similar with the calculation between layers in the forward propagation of the neural network algorithm. From this point of view we establish a rigorous corresponding relationship between the deep neural network (DNN) and RG. Concretely, we consider the most general fully connected network structure and real space RG of one dimensional Ising model. We prove that when the parameters of neural network achieve their optimal value, the limit of coupling constant of the output of neural network equals to the fixed point of the coupling constant in RG of one dimensional Ising model. This conclusion shows that the training process of neural network is equivalent to RG and therefore the network extract macroscopic feature from the input data just like RG.

\end{abstract}
\section{Introduction}
\hspace{1em}Deep neural networks are widely used in different area like pattern recognition, natural language processing, image synthesising, reinforcement learning, etc. But they are frequently said to be a black box model, which means that it is hard to explain why and how the model works, the calculation and the content (like the value of the parameters) in the model can not be directly understand by mankind. Therefore the interpretability is a topic of general interest among the researchers of machine learning.\\

\hspace{1em}There are many different aspects and methods to study the interpretability of deep neural networks. In the review \cite{ref1} and \cite{ref2}, they enumerated some major works on the interpretability, for example, in \cite{ref2}, they classified the interpretation methods into two classes: post-hoc interpretability analysis and ad-hoc interpretable modeling. The post-hoc methods try to explain the model after the training finished, like feature analysis \cite{ref3}, model inspection \cite{ref4}, etc. The ad-hoc interpretable modeling try to build models with good interpretability by designing particular structures. In this paper, we do the post-hoc interpretability analysis using mathematical and physical analysis, we put the neural network in a theoretical framework and interpret how it works.

\hspace{1em}In our work we consider the renormalization group. There are some other works started from this point of view. The first notable work is that in 2014, Mehta \cite{ref5} constructed a mapping between the variational renormalization group and restricted Boltzmann machines (RBM). In this paper the mapping is intuitional, but it did not involve the training process of the network. Also, it only compared with RBM but not for general deep neural networks. Then there were some experimental results based on this work, like \cite{ref6}. Other work in recent years also considered a special kind of concrete machine learning model, for example, \cite{ref7} compared renormalization group and auto-encoders through transfer learning, \cite{ref8}
compared renormalization group and PCA. In our work, we establish a rigorous corresponding relationship between RG and the training process of general DNNs. The framework we propose is also general and one can try to utilize it to explain all kinds of neural networks with different structures.

\section{Background}
\subsection{Deep neural network}
\hspace{1em}The network structure we use in this paper is the simplest fully connected neural network with multiple hidden layers. The details on the structure of the network will be introduced in latter sections. The loss function of the model we use is:
\begin{align}
L(\mathbf{W}^{(t)})=\frac{1}{2}||\hat{y}(\mathbf{W}^{(t)})-\hat{y}({\mathbf{W}}^*)||_2^2+{\lambda}||\mathbf{W}^{(t)}||^2_2.
\end{align}
Here $\mathbf{W}^{(t)}$ is the parameter at time $t$ of training in the network, $\hat{y}$ denote the output of the network and we omit the data input to the network. $\mathbf{W}^*$ is the optimal value of the network parameters. The second term is the regularization term.\\

The training algorithm we consider is simulated annealing. The continuation of this time discrete algorithm is the Langevin diffusion which satisfies the following stochastic differential equation:
\begin{align}
d\mathbf{W}^{(t)}=-{\nabla}L(\mathbf{W}^{(t)})dt+\sqrt{{\eta}_t}dB^{(t)},t{\geq}0.
\end{align}
\\Where $B^{(t)}$ is the standard Brownian motion with the same dimension with parameter $\mathbf{W}^{(t)}$, ${\eta}_t$ is a function of time $t$. For the initial condition, we choose $\mathbf{W}^{(0)}$ obeys a fixed distribution $P_0(\mathbf{W}^{(0)})$.\\

Holley and Stroock et al \cite{ref9} \cite{ref10} discussed the convergence of this algorithm, in our analysis we use their conclusion, similar with theorem 2.2 in \cite{ref10}, we have:
\paragraph{Lemma 1}
Suppose $\mathbf{W}^{(t)}$ satisfies equation (2), $\mathbf{W}^*=0$. Then there exists a choice of the concrete form of ${\eta_t}$, under which we have for ${\delta}>0$:
\begin{align}
\mathbb{P}(L(\mathbf{W}^{(t)}){\geq}{\delta}){\leq}{\epsilon}({\delta},t).
\end{align}
Where ${\epsilon}({\delta},t){\rightarrow}0$ when $t{\rightarrow}{\infty}$.

This result shows that in this algorithm $L(\mathbf{W}^{(t)})$ will achieve its minimal value $0$ with probability $1$ when $t{\rightarrow}{\infty}$.

\paragraph{Remark 1}
In simulation annealing algorithm, in order to ensure the existence of the stationary distribution of equation (1) when ${\eta_t}$ fixed as a constant ${\eta}$:
\begin{align}
\mathbb{P}(\mathbf{W})=\frac{1}{Z}e^{-\frac{2S}{{\eta}}L(\mathbf{W})},Z={\int}e^{-\frac{2S}{{\eta}}L(\mathbf{W})}d\mathbf{W}<{\infty}{\Rightarrow}{\lim_{\mathbf{W}{\rightarrow}{\infty}}}L(\mathbf{W})={\infty}.
\end{align}
This is the reason why we consider the regularization term in our loss function.

\subsection{Renormalization group}
\hspace{1em}The renormalization group is a method to study the critical phenomenon. The idea of it was first proposed by Kadanoff \cite{ref11}, and then was developed by Wilson et al \cite{ref12} \cite{ref13} \cite{ref14}. The method of real space renormalization group is to integrate out the variable in small length scale in order to make the fundamental part of the system, like the spin in Ising model \cite{ref15}, to form an equivalent larger "block". At the same time some significant properties of the system stay invariant when we repeat this step. More concretely, in 1-dim Ising model, this step is the coarse graining process on the one dimensional lattice. We have:

\paragraph{Lemma 2}
Suppose the probability distribution and the partition function of 1 dimensional Ising model with periodic boundary condition is:
\begin{align}
\mathbb{P}(x_1,x_2,\cdot\cdot{\cdot},x_N)&=\frac{1}{Z(J)}e^{J\sum_{i=1}^{N}x_ix_{i+1}}, x_i{\in}\{{\pm}1\},i=1,2,\cdot\cdot{\cdot},N, x_{N+1}{\triangleq}x_1,\\Z(J)&=\sum_{\{x_i\}={\pm}1}e^{J\sum_{i=1}^{N}x_ix_{i+1}}. 
\end{align}
Where constant $J>0$ is the coupling constant of the model. If we take the following coarse graining process in each step (here we suppose $N$ to be even): 
\begin{align}
{\tilde{x}}_i=x_{2i},i=1,2,\cdot\cdot{\cdot},\frac{N}{2},
\end{align}
which means we sum out the spins on the odd number position. Repeating the coarse graining process on the model, in each step keep the value of the partition function and the structure of the Hamiltonian to be invariant:
\begin{align}
Z(J)=\sum_{\{x_i\}={\pm}1}e^{J\sum_{i=1}^{N}x_ix_{i+1}}=\sum_{\{{\tilde{x}}_j\}={\pm}1}e^{J_0+{\tilde{J}}\sum_{j=1}^{\frac{N}{2}}{\tilde{x}}_i{\tilde{x}}_{i+1}}.
\end{align}
Where $J_0$ is a non-zero constant, it must be introduced to keep the equation established. Then
\begin{align}
J_0=\frac{1}{2}(log(4cosh(2J)),{\tilde{J}}=\frac{1}{2}log(cosh(2J)).
\end{align}
From this equation, the stable fixed point of $J$ is $J^*=0$. This is the critical point of the coupling constant of the system, which is a macroscopic property of the model.\\

In our work we will take 1-dim Ising model as the input of the neural network and prove that after the training of network, the corresponding coupling constant of the network output, which we will define latter, also achieves the critical point $J^*=0$. 

\section{A simple case}
\subsection{Introduction of the model}
\hspace{1em}First, we consider a simple neural network model with no hidden layers. The structure of the model is in figure 1. 
\begin{figure}[H]
	\centering
	\includegraphics[width=0.4\textwidth]{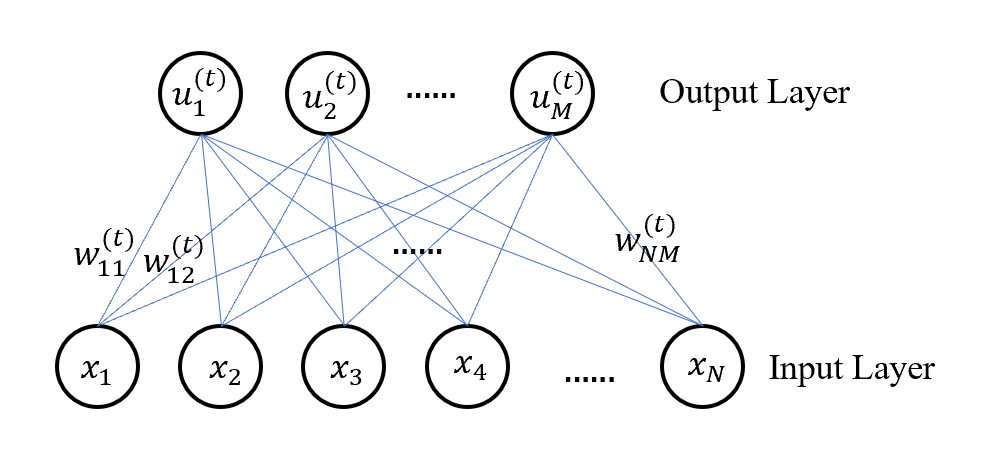}\\
	\caption{Network Structure}\label{structure}
\end{figure}
The input of the neural network is 1 dimensional Ising model, which has $N$ spins take value in $\{\pm1\}$ and obey the distribution (5). The output layer has $M$ neurone, the values $\{u_j^{(t)}\}_{j=1}^{M}$ at time $t$ of which come from the linear combination of the input under the action of the sigmoid activation function:
\begin{align}u_j^{(t)}=f(x_1w_{1j}^{(t)}+x_2w_{2j}^{(t)}+\cdot\cdot\cdot+x_Nw_{Nj}^{(t)}), f(x)=\frac{1}{1+e^{-x}}.\end{align}
$w_{ij}^{(t)}$ is the weight from $i$-th input $x_i$ to $j$-th output $u_j^{(t)}$ at time $t$. Here we suppose the optimal value of the weights of the network are $w_{ij}^{*}=0,i=1,\cdot\cdot{\cdot}N,j=1,\cdot\cdot{\cdot}M.$ Therefore according to (1) the loss function of the model equals to:\\
\begin{align}
L(\mathbf{W}^{(t)})&=\frac{1}{2}\sum_{\{x_i\}={\pm}1}[\sum_{j=1}^{M}(f(x_1w_{1j}^{(t)}+x_2w_{2j}^{(t)}+\cdot\cdot\cdot+x_Nw_{Nj}^{(t)})-f(x_1w_{1j}^{*}+x_2w_{2j}^{*}+\cdot\cdot\cdot+x_Nw_{Nj}^{*}))^2]+{\lambda}\sum_{i=1}^{N}\sum_{j=1}^{M}(w_{ij}^{(t)})^2\\&=\frac{1}{2}\sum_{\{x_i\}={\pm}1}[\sum_{j=1}^{M}(f(x_1w_{1j}^{(t)}+x_2w_{2j}^{(t)}+\cdot\cdot\cdot+x_Nw_{Nj}^{(t)})-\frac{1}{2})^2]+{\lambda}\sum_{i=1}^{N}\sum_{j=1}^{M}(w_{ij}^{(t)})^2.
\end{align}
$L$ then has a unique minimal point $\mathbf{W}^{(t)}=0$. $\mathbf{W}^{(t)}$ satisfies the stochastic differential equation:
\begin{align}
d\mathbf{W}^{(t)}=-{\nabla}L(\mathbf{W}^{(t)})dt+\sqrt{{\eta}_t}dB^{(t)}.
\end{align}
\\Where $B^{(t)}$ is the $(N*M)$-dim standard Brownian motion. We choose ${\eta_t}$ properly according to lemma 1. For the initial condition, we choose $\mathbf{W}^{(0)}$ obeys a fixed distribution $P_0(\mathbf{W}^{(0)})$.

\subsection{Method of establishing the corresponding relationship}
\hspace{1em}Now we establish the corresponding relationship between the neural network model mentioned above and the real space renormalization group of the 1-dim Ising model. Concretely, we keep the following equation to be satisfied on $t{\in}[0, +{\infty})$:
\begin{align}
Z(J){\equiv}e^{Ng(t)}\mathbb{E}[\sum_{\{x_i\}={\pm}1}e^{J_t\sum_{j=1}^{M}f(x_1w_{1j}^{(t)}+x_2w_{2j}^{(t)}+\cdot\cdot\cdot+x_Nw_{Nj}^{(t)})f(x_1w_{1,j+1}^{(t)}+x_2w_{2,j+1}^{(t)}+\cdot\cdot\cdot+x_Nw_{N,j+1}^{(t)})}].
\end{align}
\hspace{1em}In this equation, $Z(J)$ is the partition function (6) of the 1-dim Ising model as the input of our model and right hand side is the partition function we define for the output layer. $g(t)$ is a function of $t$ with no singularity, which is correspond to $J_0$ in lemma 2. $J_t$ is the value which playes the role similar with the coupling constant $J$ of Ising model. $J_t$ is a function of $t$, which represents the renormalization process done by the training of neural network over time. Here we still use the periodic boundary condition: $f(x_1w_{1,M+1}^{(t)}+x_2w_{2,M+1}^{(t)}+\cdot\cdot\cdot+x_Nw_{N,M+1}^{(t)})=f(x_1w_{11}^{(t)}+x_2w_{21}^{(t)}+\cdot\cdot\cdot+x_Nw_{N1}^{(t)}).$
For the initial value, we have:
\begin{align}
Z(J)=\mathbb{E}_{P_0(\mathbf{W}^{(0)})}[\sum_{\{x_i\}={\pm}1}e^{J_0\sum_{j=1}^{M}f(x_1w_{1j}^{(0)}+x_2w_{2j}^{(0)}+\cdot\cdot\cdot+x_Nw_{Nj}^{(0)})f(x_1w_{1,j+1}^{(0)}+x_2w_{2,j+1}^{(0)}+\cdot\cdot\cdot+x_Nw_{N,j+1}^{(0)})}].
\end{align}
So for $P_0(\mathbf{W}^{(0)})$ we should choose a distribution such that value $J_0$ exists. And in order to be consistent with the renormalization group of 1-dim Ising model, the initial value $g(0)$ of fuction $g$ should be $0$.\\

The criterion we obey to establish this equation are: (1) The value of the partition function of the output layer is equal to the first layer, and it remains unchanged when time $t$ varies from $0$ to infinity. (2) The structure of the Hamiltonian remains unchanged when time $t$ varies from $0$ to infinity. Here the coarse graining is done by the neural network and at time step $t$, the result of coarse graining is the value of output neurone $\{u_j^{(t)}\}_{j=1}^{M}$. So the two criterion are the same as the criterion in real space renormalization group of 1-dim Ising model, which makes our corresponding relationship rigorous. Our result is the following theorem:\\
\paragraph{Theorem 1}
With above model sturcture and algorithm, there exists a function $g(t){\in}C^{\infty}$, such that when equation (14) holds for $t=0$ and equation (15) holds for $t{\in}[0,\infty)$, we have $\lim_{t{\rightarrow}{\infty}}J_t=J^{*}=0$. $J^{*}$ denotes the fixed point of $J$ in real space renormalization group of 1-dim Ising model.

\paragraph{Proof}According to lemma 1, we have for ${\delta}>0$:
\begin{align}
\mathbb{P}(L(\mathbf{W}^{(t)}){\geq}{\delta}){\leq}{\epsilon}_1({\delta},t).
\end{align}
Where ${\epsilon}_1{\geq}0$ is a function of ${\delta}$ and $t$, which satisfies ${\lim_{t{\rightarrow}{\infty}}}{\epsilon}_1({\delta},t)=0$. If we assume
\begin{align}
 {\exists}j{\in}\{1,\cdot\cdot\cdot,M\}, s.t. ||\mathbf{W}_{j}^{(t)}||_{1}{\triangleq}|w_{1j}^{(t)}|+|w_{2j}^{(t)}|+\cdot\cdot\cdot+|w_{Nj}^{(t)}|{\geq}{\delta}_2,{\delta}_2>0,
\end{align}
then we have:
\begin{align}
L(\mathbf{W}^{(t)})&{\geq}\frac{1}{2}\sum_{\{x_i\}={\pm}1}(f(x_1w_{1j}^{(t)}+x_2w_{2j}^{(t)}+\cdot\cdot\cdot+x_Nw_{Nj}^{(t)})-\frac{1}{2})^2\\
&{\geq}\frac{1}{2}(f(|w_{1j}^{(t)}|+|w_{2j}^{(t)}|+\cdot\cdot\cdot+|w_{Nj}^{(t)}|)-\frac{1}{2})^2.
\end{align}
The second inequality comes from that in the summation, values of $\{x_i\}_{i=1}^{N}$ are taken from all of the combinations in $\{{\pm}1\}^{N}$, so there always exist one term of $f(x_1w_{1j}^{(t)}+x_2w_{2j}^{(t)}+\cdot\cdot\cdot+x_Nw_{Nj}^{(t)})$ to be equal to $f(|w_{1j}^{(t)}|+|w_{2j}^{(t)}|+\cdot\cdot\cdot+|w_{Nj}^{(t)}|)$. For the sigmoid function $f(x)$, when $x{\geq}{\delta_2}>0$, $(f(x)-\frac{1}{2})^2{\geq}(f({\delta_2})-\frac{1}{2})^2$. Therefore,

\begin{align}
L(\mathbf{W}^{(t)}){\geq}\frac{1}{2}(f({\delta_2})-\frac{1}{2})^2{\triangleq}{\delta_1}.
\end{align}
According to (16), we have:

\begin{align}
\mathbb{P}(A){\triangleq}\mathbb{P}({\exists}j{\in}\{1,\cdot\cdot\cdot,M\}, s.t. ||\mathbf{W}_{j}^{(t)}||_{1}{\geq}{\delta}_2){\leq}{\epsilon}_1({\delta_1},t),\\
\mathbb{P}(A^c)=\mathbb{P}({\forall}j{\in}\{1,\cdot\cdot\cdot,M\},||\mathbf{W}_{j}^{(t)}||_{1}<{\delta}_2)>1-{\epsilon}_1({\delta_1},t).
\end{align}
Let $F(\mathbf{W}^{(t)},J_t)=e^{J_t\sum_{j=1}^{M}f(x_1w_{1j}^{(t)}+x_2w_{2j}^{(t)}+\cdot\cdot\cdot+x_Nw_{Nj}^{(t)})f(x_1w_{1,j+1}^{(t)}+x_2w_{2,j+1}^{(t)}+\cdot\cdot\cdot+x_Nw_{N,j+1}^{(t)})},$ we obtain
\begin{align}
\frac{Z(J)}{e^{Ng(t)}}&=\mathbb{E}[F(\mathbf{W}^{(t)},J_t)]=\mathbb{E}[F(\mathbf{W}^{(t)},J_t)\mathbb{I}_{A}+F(\mathbf{W}^{(t)},J_t)\mathbb{I}_{A^c}]\\&{\geq}\mathbb{E}[F(\mathbf{W}^{(t)},J_t)\mathbb{I}_{A^c}].
\end{align}
Where $\mathbb{I}_{A}$ denotes the indicator function of set $A$. On set $A^c$:
\begin{align}
F(\mathbf{W}^{(t)},J_t)&{\geq}2^Ne^{J_t\sum_{j=1}^{M}f(-\sum_{i=1}^{N}|w^{(t)}_{ij}|)f(-\sum_{i=1}^{N}|w^{(t)}_{i,j+1}|)}\\&>2^Ne^{J_tMf(-{\delta_2})^2}.
\end{align}
Then 
\begin{align}
\frac{Z(J)}{e^{Ng(t)}}&=\mathbb{E}[F(\mathbf{W}^{(t)},J_t)]>2^Ne^{J_tMf(-{\delta_2})^2}(1-{\epsilon}_1({\delta_1},t)).
\end{align}
There exists $T>0$ s.t. if $t>T$, $1-{\epsilon}_1({\delta_1},t)>0$. Above inequality is equivalent to:

\begin{align}
J_t<\frac{1}{2f(-{\delta_2})^2}log(\frac{Z(J)}{2^Ne^{Ng(t)}}\cdot\frac{1}{1-{\epsilon}_1({\delta_1},t)}),
\end{align}
when $t>T$. If we choose $g(t)$ properly, for example, $g(t)=(1-e^{-t})(\frac{1}{N}logZ(J)-log2)$, then $g(0)=0$, and
\begin{align}
\lim_{t{\rightarrow}{\infty}}\frac{1}{2f(-{\delta_2})^2}log(\frac{Z(J)}{2^Ne^{Ng(t)}}\cdot\frac{1}{1-{\epsilon}_1({\delta_1},t)})=\lim_{t{\rightarrow}{\infty}}\frac{1}{2f(-{\delta_2})^2}log(\frac{1}{1-{\epsilon}_1({\delta_1},t)})=0.
\end{align}
Since $J_t{\geq}0$, $\lim_{t{\rightarrow}{\infty}}J_t=J^*=0$. This ends the proof of Theorem 1.

According to theorem 1, when the simmulation annealing process achieves its optimal point, which happens with probability 1 when $t{\rightarrow}{\infty}$, the coupling constant $J_t$ of output of this neural network tends to $0$, exactly the same as the critical point of coupling constant in 1-dim Ising model, the input of network. On the other hand, similar with the renormalization of 1-dim Ising model, function $g(t)$, correspond to $J_0$ in lemma 2, represents the non singular part of the Hamiltonian. On the contrary,
$J_t\sum_{j=1}^{M}f(x_1w_{1j}^{(t)}+x_2w_{2j}^{(t)}+\cdot\cdot\cdot+x_Nw_{Nj}^{(t)})f(x_1w_{1,j+1}^{(t)}+x_2w_{2,j+1}^{(t)}+\cdot\cdot\cdot+x_Nw_{N,j+1}^{(t)})$ represents the singular part of the Hamiltonian correspond to ${\tilde{J}}\sum_{j=1}^{\frac{N}{2}}{\tilde{x}}_i{\tilde{x}}_{i+1}$ part in lemma 2. When $t{\rightarrow}{\infty}$, equation (14) becomes $Z(J)=2^Ne^{Ng({\infty})}$ since $J_{\infty}=0$. And we have $\frac{1}{N}log2^Ne^{Ng({\infty})}=g({\infty})+log2=\frac{1}{N}logZ(J)$,
from this we can see that the mean freedom energy per single particle remains unchange in the whole process.

\paragraph{Remark 2}In our proof of theorem 1, we do not refer to the regularization term in the loss function and just use it is greater or equal to $0$ in the inequality (18). We should mention here that if we consider the stochastic gradient descent (SGD) algorithm:

\begin{align}
d\mathbf{W}^{(t)}&=-{\nabla}L(\mathbf{W}^{(t)})dt+\sqrt{\frac{{\eta}_t}{S}}{\sigma}(\mathbf{W}^{(t)})dB^{(t)}, t{\ge}0,\\
{\sigma}(\mathbf{W}^{(t)})&{\sigma}(\mathbf{W}^{(t)})^T={\Sigma}(\mathbf{W}^{(t)})=\frac{1}{2^N}\sum_{\{x_i\}={\pm}1}[({\nabla}l(\mathbf{W}^{(t)},\{x_i\})-{\nabla}L(\mathbf{W}^{(t)}))({\nabla}l(\mathbf{W}^{(t)},\{x_i\})-{\nabla}L(\mathbf{W}^{(t)}))^T],\\
l(\mathbf{W}^{(t)},\{x_i\})&=\frac{1}{2}[\sum_{j=1}^{M}(f(x_1w_{1j}^{(t)}+x_2w_{2j}^{(t)}+\cdot\cdot\cdot+x_Nw_{Nj}^{(t)})-f(x_1w_{1j}^{*}+x_2w_{2j}^{*}+\cdot\cdot\cdot+x_Nw_{Nj}^{*}))^2]+{\frac{{\lambda}}{2^N}}\sum_{i=1}^{N}\sum_{j=1}^{M}(w_{ij}^{(t)})^2,\\
\mathbf{W}^{(0)}&{\sim}P_0(\mathbf{W}^{(0)}).
\end{align}
Then our frame and method still work, if there exist the convergent result with form in lemma 1 for SGD. We consider simmulation annealing here since there are no such results now.

\section{The corresponding relationship with deep neural network}

\hspace{1em}In this section we consider the general case: to establish the relationship between the renormalization group and the deep neural network. The network structure we consider is still the full connect neural network but with $L$ hidden layers, the number of neurone in $l$-th hidden layer is $H_{l}$. The weights parameters of the $l$-th layer in the network are $\mathbf{W}^{(l)}=(w^{(l)}_{ij}),  i=1,\cdot\cdot\cdot,H_{l-1}, j=1,\cdot\cdot\cdot,H_{l}, l=1,\cdot\cdot\cdot,L+1$. The bias parameter of the $l$-th layer is $\mathbf{B}^{(l)}=(b^{(l)}_{k}), k=1,\cdot\cdot\cdot,H_{l}, l=1,\cdot\cdot\cdot,L+1.$ Here for convenience we omit the label of time $t$ but we emphasize that $\mathbf{W}^{(l)}$ and $\mathbf{B}^{(l)}$ depends on $t$ through the training process. The input of network is still 1-dim Ising model represented by $\{x_i\}, i=1,\cdot\cdot\cdot,N.$ We denote the output by ${\widehat{y}}_{k}^{(t)}, k=1,\cdot\cdot\cdot,M.$ The activation function $f$ is the sigmoid function. The expression of values in the network are ($u^{(l)}_k$ denote the value of the $k$-th neuron in $l$-th layer):

\begin{align}
u^{(1)}_{k}&=f(x_1w^{(1)}_{1k}+x_2w^{(1)}_{2k}+\cdot\cdot\cdot+x_Nw^{(1)}_{Nk}+b^{(1)}_{k}), k=1,\cdot\cdot\cdot,H_1.\\
u^{(l)}_{k}&=f(u^{(l-1)}_1w^{(l)}_{1k}+u^{(l-1)}_2w^{(l)}_{2k}+\cdot\cdot\cdot+u^{(l-1)}_{H_{l-1}}w^{(l)}_{H_{l-1}k}+b^{(1)}_{k}), l=2,\cdot\cdot\cdot,L, k=1,\cdot\cdot\cdot,H_l.\\
{\widehat{y}}_{k}^{(t)}&=f(u^{(L)}_1w^{(L+1)}_{1k}+u^{(L)}_2w^{(L+1)}_{2k}+\cdot\cdot\cdot+u^{(L)}_{H_{L}}w^{(L+1)}_{H_{L}k}+b^{(L+1)}_{k}),  k=1,\cdot\cdot\cdot,M.
\end{align}

The loss function of this network is (we still suppose the optimal value of the parameters $\mathbf{W}^{*}=\mathbf{B}^{*}=0$):

\begin{align}
L(\mathbf{W},\mathbf{B})=&\frac{1}{2}\sum_{\{x_i\}={\pm}1}[\sum_{j=1}^{M}(f(u^{(L)}_1w^{(L+1)}_{1j}+u^{(L)}_2w^{(L+1)}_{2j}+\cdot\cdot\cdot+u^{(L)}_{H_{L}}w^{(L+1)}_{H_{L}j}+b^{(L+1)}_{j})-\frac{1}{2})^2]\\&+{\lambda}(\sum_{l=1}^{L+1}\sum_{i=1}^{H_{l-1}}\sum_{j=1}^{H_{l}}(w_{ij}^{(l)})^2+\sum_{l=1}^{L+1}\sum_{i=1}^{H_{l}}(b^{(l)}_{i})^2).
\end{align}
 The alogrithm we use is still simmulation annealing, So the stochastic differential equation $\mathbf{{\Theta}}^{(t)}{\triangleq}(\mathbf{W}^{(t)},\mathbf{B}^{(t)})$ satisfies is:
 \begin{align}
 d\mathbf{{\Theta}}^{(t)}=-{\nabla}L(\mathbf{{\Theta}}^{(t)})dt+\sqrt{{\eta}_t}dB^{(t)}.
 \end{align}

Here we still choose ${\eta_t}$ according to lemma 1. As in equation (14), we write:

\begin{align}
Z(J){\equiv}e^{Ng(t)}\mathbb{E}[\sum_{\{x_i\}={\pm}1}e^{J_t\sum_{j=1}^{M}f(u^{(L)}_1w^{(L+1)}_{1j}+u^{(L)}_2w^{(L+1)}_{2j}+\cdot\cdot\cdot+u^{(L)}_{H_{L}}w^{(L+1)}_{H_{L}j}+b^{(L+1)}_{j})f(u^{(L)}_1w^{(L+1)}_{1,j+1}+u^{(L)}_2w^{(L+1)}_{2,j+1}+\cdot\cdot\cdot+u^{(L)}_{H_{L}}w^{(L+1)}_{H_{L},j+1}+b^{(L+1)}_{j+1})}].
\end{align}
The boundary condition and the initial condition is the same as those in section 3:

\begin{align}
f(u^{(L)}_1&w^{(L+1)}_{1,M+1}+u^{(L)}_2w^{(L+1)}_{2,M+1}+\cdot\cdot\cdot+u^{(L)}_{H_{L}}w^{(L+1)}_{H_{L},M+1}+b^{(L+1)}_{M+1})=f(u^{(L)}_1w^{(L+1)}_{11}+u^{(L)}_2w^{(L+1)}_{21}+\cdot\cdot\cdot+u^{(L)}_{H_{L}}w^{(L+1)}_{H_{L}1}+b^{(L+1)}_{1}),\\
Z(J)=&\mathbb{E}_{P_0(\mathbf{W}^{(0)},\mathbf{B}^{(0)})}[\sum_{\{x_i\}={\pm}1}exp\{J_0\sum_{j=1}^{M}f(u^{(L),(0)}_1w^{(L+1),(0)}_{1j}+\cdot\cdot\cdot+u^{(L),(0)}_{H_{L}}w^{(L+1),(0)}_{H_{L}j}+b^{(L+1),(0)}_{j})\\&{\cdot}f(u^{(L),(0)}_1w^{(L+1),(0)}_{1,j+1}+\cdot\cdot\cdot+u^{(L),(0)}_{H_{L}}w^{(L+1),(0)}_{H_{L},j+1}+b^{(L+1),(0)}_{j+1})\}].
\end{align}
Where $P_0(\mathbf{W}^{(0)},\mathbf{B}^{(0)})$ is a distribution of initial value of the parameters, under which $J_0$ exists. Just like section 3, we have following theorem:

\paragraph{Theorem 2}
With above model sturcture and algorithm, there exists a function $g(t){\in}C^{\infty}$, such that when equation (42) holds for $t=0$ and equation (40) holds for $t{\in}[0,\infty)$, we have $\lim_{t{\rightarrow}{\infty}}J_t=J^{*}=0$. $J^{*}$ denotes the fixed point of $J$ in real space renormalization group of 1-dim Ising model.

\paragraph{Proof}The method is similar with the proof of theorem 1. First we have:
\begin{align}
\mathbb{P}(L(\mathbf{{\Theta}}^{(t)}){\geq}{\delta}){\leq}{\epsilon}_1({\delta},t).
\end{align}
Where ${\epsilon}_1{\geq}0$ is a function of ${\delta}$ and $t$, which satisfies ${\lim_{t{\rightarrow}{\infty}}}{\epsilon}_1({\delta},t)=0$. Now if we assume:
\begin{align}
{\exists}j{\in}\{1,\cdot\cdot\cdot,M\}, s.t. ||W^{(L+1)}_j||_2{\geq}{\delta_1}>0, or |b^{(L+1)}_j|{\geq}{\delta_2}>0.
\end{align}
Where $W^{(L+1)}_j=(w^{(L+1)}_{1j},w^{(L+1)}_{2j},\cdot\cdot\cdot,w^{(L+1)}_{H_Lj})$. Then
\begin{align}
L(\mathbf{{\Theta}}){\geq}{\lambda}max\{{\delta_1},{\delta_2}\}{\triangleq}{\delta}.
\end{align}
Therefore
\begin{align}
\mathbb{P}(A){\triangleq}\mathbb{P}({\exists}j{\in}\{1,\cdot\cdot\cdot,M\}, s.t. ||W^{(L+1)}_j||_2{\geq}{\delta_1}>0, or |b^{(L+1)}_j|{\geq}{\delta_2}>0){\leq}{\epsilon}_1({\delta_1},t),\\
\mathbb{P}(A^c)=\mathbb{P}({\forall}j{\in}\{1,\cdot\cdot\cdot,M\},||W^{(L+1)}_j||_2<{\delta_1}and |b^{(L+1)}_j|<{\delta_2})>1-{\epsilon}_1({\delta_1},t).
\end{align}
Define the expression in the expectation of equation (40) to be $F(J_t,{\Theta}^{(t)})$, the inequalities about probability will yield:
\begin{align}
\frac{Z(J)}{e^{Ng(t)}}&=\mathbb{E}[F(J_t,{\Theta}^{(t)})]=\mathbb{E}[F(J_t,{\Theta}^{(t)})\mathbb{I}_{A}+F(J_t,{\Theta}^{(t)})\mathbb{I}_{A^c}]\\&{\geq}\mathbb{E}[F(J_t,{\Theta}^{(t)})\mathbb{I}_{A^c}]>2^Ne^{MJ_tf(-{\delta_1}H_L-{\delta_2})^2}(1-{\epsilon}_1({\delta_1},t)).
\end{align}
The last inequality comes from: on set $A^c$, ${\forall}j,$ since $(w^{(L+1)}_{1j})^2+\cdot\cdot\cdot+(w^{(L+1)}_{H_Lj})^2<{\delta_1}^2$, we have ${\forall}i, (w^{(L+1)}_{ij})^2<{\delta_1}^2, w^{(L+1)}_{ij}>-{\delta_1}.$ and $b^{(L+1)}_j>-{\delta_2}$, $f$ is monotonically increasing,
\begin{align}
f(u^{(L)}_1w^{(L+1)}_{1j}+\cdot\cdot\cdot+u^{(L)}_{H_L}w^{(L+1)}_{H_Lj}+b^{(L+1)}_j)>f(-{\delta_1}H_L-{\delta_2}), {\forall}j{\in}\{1,\cdot\cdot\cdot,M\}.
\end{align}
So we can get the bound of $J_t$ ($T$ is the same as in section 3):
\begin{align}
J_t<\frac{1}{Mf(-{\delta_1}H_L-{\delta_2})^2}log(\frac{Z(J)}{2^Ne^{Ng(t)}(1-{\epsilon}_1({\delta_1},t))}),t>T.
\end{align}
When we choose $g(t)=(1-e^{-t})(\frac{1}{N}logZ(J)-log2)$ then $\lim_{t{\rightarrow}{\infty}}J_t=J^*=0$. This finishes the proof of Theorem 2.\\
\hspace{1em}As a result, theorem 2 shows that for deep neural networks, the training process achieves the same goal as the renormalization group on 1-dim Ising model. And this is the reason why the network can abstract the macroscopic scale features of the input data.

\section{Conclusion and future work}
\hspace{1em}In this paper we proved the equivalence of the training process of a neural network and renormalization group on a concrete example. This explains that the deep neural network performs well on many situations because it just does the same thing as the renormalization group, which makes it has the capacity to extract the macroscopic features of the input data. And it should be emphasized that the method we propose is a general framework to establish the relationship between neural network and RG. In the future we can try to use this framework on other type of deep learning model or input data to get the interpretability of more models.

\end{document}